\pdfoutput=1
\documentclass[aps,prd,amsmath,floats,floatfix, twocolumn,
superscriptaddress,nofootinbib,showpacs,longbibliography]{revtex4-1}
\usepackage[T1]{fontenc}
\usepackage[utf8]{inputenc}
\usepackage{txfonts}
\usepackage{verbatim}
\usepackage[dvipsnames, usenames]{xcolor}
\definecolor{linkcolor}{rgb}{0.0,0.3,0.5}
\usepackage[hypertexnames=false, unicode, colorlinks=true, linkcolor=linkcolor,
citecolor=linkcolor, filecolor=linkcolor,urlcolor=linkcolor,
pdfusetitle]{hyperref}
\usepackage[all]{hypcap}
\usepackage{graphicx}
\usepackage{xspace}
\usepackage{amssymb}
\usepackage{microtype}
\usepackage{subfigure}

\usepackage{url}

\usepackage[english]{babel}
\usepackage{blindtext}

\usepackage[normalem]{ulem} 
\usepackage{bm} 
\usepackage{mathrsfs}
\usepackage{xspace} 

\DeclareMathAlphabet{\mathpzc}{OT1}{pzc}{m}{it}

\usepackage[nomessages]{fp}

\newcommand{\sk}[1]{}

\begin{document}
\title{Data-driven extraction and phenomenology of eccentric harmonics in \\eccentric spinning binary black hole mergers}
\newcommand{\KITP}{\affiliation{Kavli Institute for Theoretical Physics, University of California Santa Barbara, Kohn Hall, Lagoon Rd, Santa Barbara, CA 93106}} 
\newcommand{\TAPIR}{\affiliation{Theoretical AstroPhysics Including Relativity and Cosmology, California Institute of Technology, Pasadena, California, USA}}
\author{Tousif Islam}
\email{tousifislam@ucsb.edu}
\KITP
\author{Tejaswi Venumadhav}
\affiliation{\mbox{Department of Physics, University of California at Santa Barbara, Santa Barbara, CA 93106, USA}}
\affiliation{\mbox{International Centre for Theoretical Sciences, Tata Institute of Fundamental Research, Bangalore 560089, India}}
\author{Ajit Kumar Mehta}
\affiliation{\mbox{Department of Physics, University of California at Santa Barbara, Santa Barbara, CA 93106, USA}}
\author{Digvijay Wadekar}
\affiliation{\mbox{Department of Physics and Astronomy, Johns Hopkins University,
3400 N. Charles Street, Baltimore, Maryland, 21218, USA}}
\affiliation{\mbox{School of Natural Sciences, Institute for Advanced Study, 1 Einstein Drive, Princeton, NJ 08540, USA}}
\author{Javier Roulet}
\affiliation{\mbox{TAPIR, Walter Burke Institute for Theoretical Physics, California Institute of Technology, Pasadena, CA 91125, USA}}
\author{Isha Anantpurkar}
\affiliation{\mbox{Department of Physics, University of California at Santa Barbara, Santa Barbara, CA 93106, USA}}
\author{Jonathan Mushkin}
\affiliation{\mbox{Department of Particle Physics \& Astrophysics, Weizmann Institute of Science, Rehovot 76100, Israel}}
\author{Barak Zackay}
\affiliation{\mbox{Department of Particle Physics \& Astrophysics, Weizmann Institute of Science, Rehovot 76100, Israel}}
\author{Matias Zaldarriaga}
\affiliation{\mbox{School of Natural Sciences, Institute for Advanced Study, 1 Einstein Drive, Princeton, NJ 08540, USA}}

\hypersetup{pdfauthor={Islam et al.}}

\date{\today}

\begin{abstract}
Newtonian and post-Newtonian (PN) calculations indicate that the phenomenology of eccentric binary black hole (BBH) merger waveforms is significantly more complex than that of their quasi-circular counterparts. Each spherical harmonic mode of the radiation can be further decomposed into several eccentricity-induced components, referred to as \textit{eccentric harmonics}. Unlike the (cumulative) spherical harmonic modes, these constituent eccentric harmonics exhibit monotonically time-varying amplitudes and frequencies. However, these eccentric harmonics are not directly accessible in numerical relativity (NR) simulations or current eccentric waveform models. Using the recently developed data-driven framework \texttt{gwMiner}~\footnote{\href{https://github.com/tousifislam/gwMiner}{https://github.com/tousifislam/gwMiner}}—which combines singular value decomposition, input from post-Newtonian theory, and signal processing techniques—we extract eccentric harmonics from eccentric, aligned-spin waveforms for six different spherical harmonic modes: $(\ell, m) = (2,1), (2,2), (3,2), (3,3), (4,3), (4,4)$. We demonstrate that the phase (frequency) of each eccentric harmonic takes the form $j\,\phi_{\ell,m,\lambda} + \phi_{\ell,m,\mathrm{ecc}}$ ($j\,f_{\ell,m,\lambda} + f_{\ell,m,\mathrm{ecc}}$), where $\phi_{\ell,m,\lambda}$ ($f_{\ell,m,\lambda}$) corresponds to the secular orbital phase (frequency), and $\phi_{\ell,m,\mathrm{ecc}}$ ($f_{\ell,m,\mathrm{ecc}}$) is an additional contribution that depends solely on the eccentricity. We further find that $\phi_{\ell,m,\lambda}$ is the same across different spherical harmonic modes $(\ell, m)$, whereas the eccentric correction term $\phi_{\ell,m,\mathrm{ecc}}$ scales with $\ell$. Using effective-one-body dynamics, we further show that $\phi_{\ell,m,\lambda}$ is nothing but the relativistic anomaly and $\phi_{\ell,m,\mathrm{ecc}}$ is related to the precession advances. Overall, our eccentric harmonic framework can be useful in quick and efficient searches and parameter estimation of eccentric asymmetric binaries.
\end{abstract}

\maketitle

\section{Introduction}
\label{sec:intro}
Eccentric binary black holes (BBHs) are among the most anticipated sources of gravitational waves (GWs). In particular, BBHs formed in dense globular clusters or galactic nuclei may retain significant eccentricity even when they enter the sensitivity band of current GW detectors~\cite{Rodriguez:2017pec,Rodriguez:2018pss,Samsing:2017xmd,Zevin:2018kzq,Zevin:2021rtf,Samsing:2020tda}. However, none of the BBHs detected by the LIGO–Virgo–KAGRA Collaboration or other independent groups have been conclusively demonstrated to be eccentric~\cite{Harry:2010zz,VIRGO:2014yos,KAGRA:2020tym,LIGOScientific:2018mvr,LIGOScientific:2020ibl,LIGOScientific:2021usb,LIGOScientific:2021djp}. Some recent analyses, however, have explored an eccentric interpretation of selected events~\cite{Romero-Shaw:2020thy,Gayathri:2020coq,Gamba:2021gap,Ramos-Buades:2023yhy,Gupte:2024jfe,Morras:2025xfu,romeroshaw2025gw200208222617eccentricblackholebinary,Planas:2025jny}.
There are significant multi-directional efforts in the community to understand the phenomenology and modelling of eccentric BBHs. In the past several years, multiple eccentric numerical relativity (NR) simulations have been performed~\cite{Scheel:2025jct,Mroue:2010re, Healy:2017zqj,Buonanno:2006ui,Husa:2007rh,Ramos-Buades:2018azo,Ramos-Buades:2019uvh,Purrer:2012wy,Bonino:2024xrv,Ramos-Buades:2022lgf}. New theoretical calculations based on post-Newtonian (PN) approximations have been published. These simulations, along with the theoretical calculations, have helped state-of-the-art waveform models—such as effective-one-body (EOB) models~\cite{Tiwari:2019jtz, Huerta:2014eca, Moore:2016qxz, Damour:2004bz, Konigsdorffer:2006zt, Memmesheimer:2004cv, Cho:2021oai,Hinderer:2017jcs,Cao:2017ndf,Chiaramello:2020ehz,Albanesi:2023bgi,Albanesi:2022xge,Riemenschneider:2021ppj,Chiaramello:2020ehz,Ramos-Buades:2021adz,Liu:2023ldr,Huerta:2016rwp,Huerta:2017kez,Joshi:2022ocr,Wang:2023ueg,Carullo:2023kvj,Nagar:2021gss,Tanay:2016zog,Gamboa:2024hli,Morras:2025nbp,Morras:2025nlp}, phenomenological models~\cite{Hinder:2017sxy,Planas:2025feq,Chattaraj:2022tay,Setyawati:2021gom,Paul:2024ujx,Manna:2024ycx}, and NR surrogates~\cite{Islam:2021mha}—begin to incorporate eccentricity. Another important set of works has attempted to standardize the definition of eccentricity across different models by providing measures of eccentricity based on waveform observables~\cite{Mroue:2010re,Mora:2002gf,Ramos-Buades:2021adz,Shaikh:2023ypz,Boschini:2024scu,Islam:2025oiv,Shaikh:2025tae}. There are also efforts to simplify the features induced by eccentricity in several radiative quantities~\cite{Islam:2021mha,Islam:2024rhm}. However, the problem still requires further simplification.

\begin{figure*}
\includegraphics[width=\textwidth]{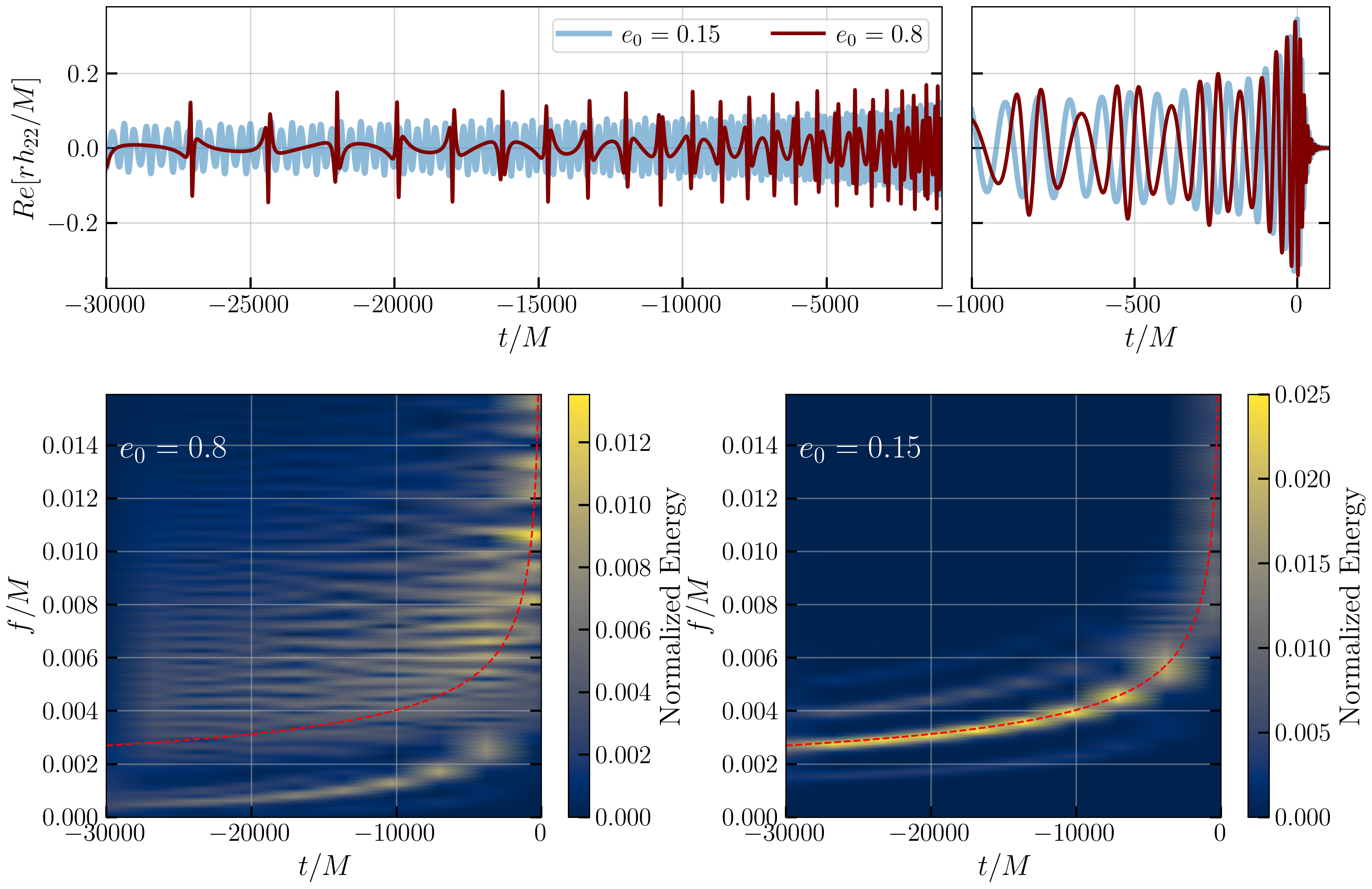}
\caption{\textit{Upper panel:} We show the real part of the quadrupolar mode gravitational waveform generated from two spinning eccentric binaries characterized by $[q,\chi_1,\chi_2,e_{\rm ref},\zeta_{\rm ref}]=[2,0.4,-0.2,0.8,\pi/2]$ (maroon line), and $[q,\chi_1,\chi_2,e_{\rm ref},\zeta_{\rm ref}]=[2,0.4,-0.2,0.15,\pi/2]$ (blue line) using \texttt{SEOBNRv5EHM} model. The reference time (where eccentricity $e_{\rm ref}$ and relativistic anomaly $\zeta_{\rm ref}$ are defined) is chosen to be $t = -30000M$, while $t=0$ denotes the time of merger. Furthermore, we present a time--frequency spectrogram obtained from the STFT in the \textit{lower panels}. For comparison, we show the circular time-frequency track in red. In each case, the full signal is a combination of multiple eccentric harmonics corresponding to different time--frequency tracks. Details are in Section~\ref{sec:time-frequency}.}
\label{fig:spectrogram}
\end{figure*}

A recent approach~\cite{Patterson:2024vbo,Islam:2025rjl} has decomposed eccentric spherical harmonic modes into several sub-harmonics, referred to as \textit{eccentric harmonics}. Although these harmonics are predicted by Newtonian calculations~\cite{Yunes:2009yz} and certain PN formalisms~\cite{VanDenBroeck:2006qu,Arun:2007qv,Seto:2001pg,2012ApJ74537V}, they are not readily accessible in existing waveform models. Even if these harmonics can be identified within existing PN expressions, extending these calculations through merger and ringdown is challenging. These sub-harmonics are all monotonic functions of time or frequency, unlike the full spherical harmonic mode~\cite{Yunes:2009yz}, making them a promising choice for template-bank construction and rapid source characterization. Their simplicity also makes them suitable for decomposition in building reduced-order surrogates based on NR or EOB waveforms. Time–frequency tracks of these harmonics have recently been explored in Refs.~\cite{Hegde:2023yoz,Bose:2021pcw}. Ref.~\cite{Patterson:2024vbo} provides an approximate framework, based on intuition from Newtonian expectations, for extracting the eccentric harmonics from a given eccentric waveform model. Ref.~\cite{Islam:2025rjl} (hereafter PAPER-I) builds on this idea and presents a robust, data-driven, PN-guided method that combines techniques such as singular value decomposition (SVD) and heterodyning to extract the eccentric harmonics from the quadrupolar mode of non-spinning eccentric BBH merger waveforms. The framework of PAPER-I is made available through the \texttt{gwMiner} package.

In this paper, we extend the framework presented in PAPER-I to both eccentric, spinning BBHs and higher-order spherical-harmonic modes. Furthermore, whereas Refs.~\cite{Patterson:2024vbo,Islam:2025rjl} extracted eccentric harmonics from the \texttt{TEOBResumS}~\cite{Chiaramello:2020ehz} EOB model, here we extract them from a different EOB variant, \texttt{SEOBNRv5EHM}~\cite{Gamboa:2024hli,Gamboa:2024imd}. We then study the eccentric harmonics and present a phenomenological hierarchy of their amplitudes and phases.

\begin{figure}
\includegraphics[width=\columnwidth]{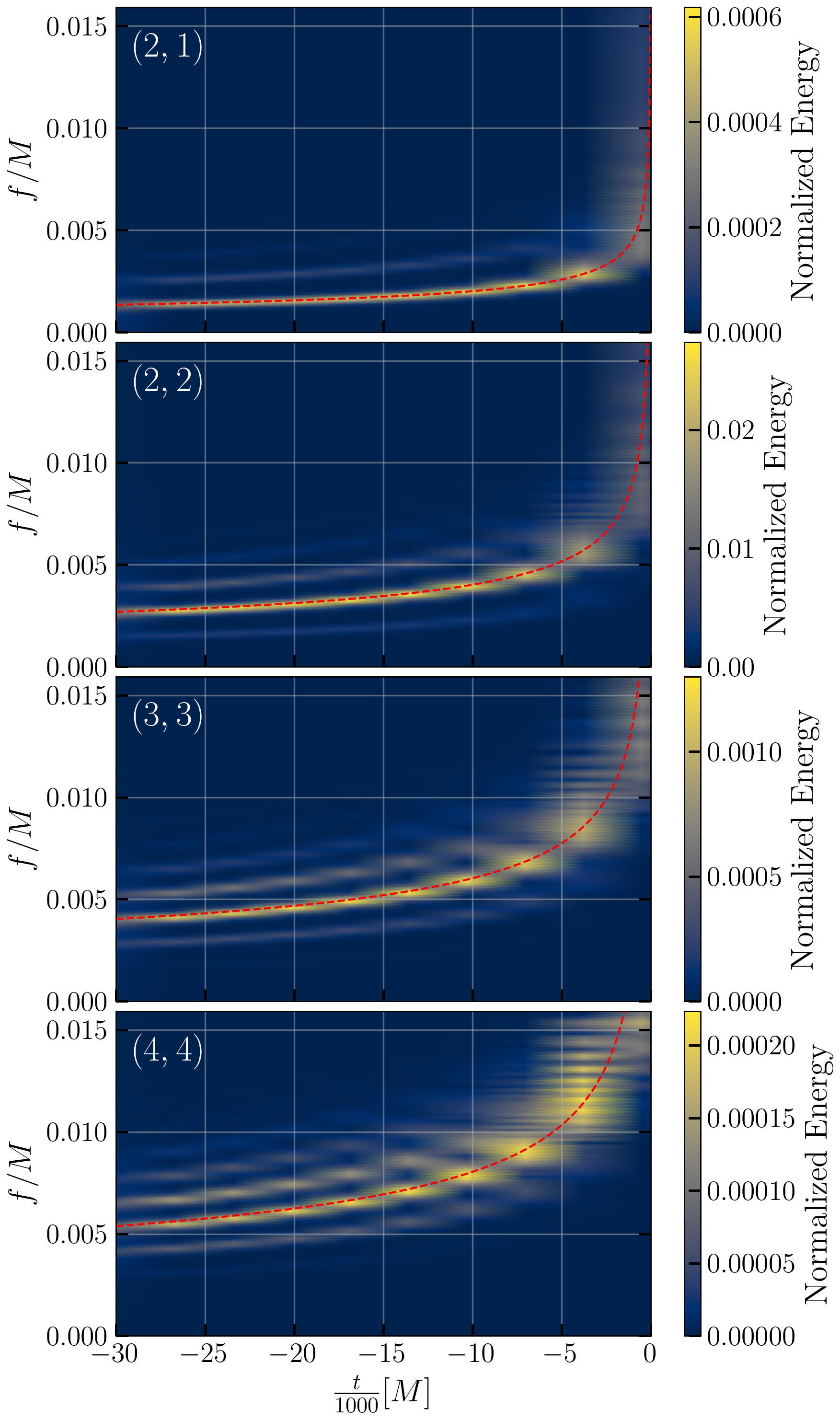}
\caption{We show the time–frequency spectrograms obtained via STFT for three representative higher-order modes—$(2,1)$ (\textit{first panel}), $(3,3)$ (\textit{third panel}), and $(4,4)$ (\textit{fourth panel})—for a spinning eccentric binary characterized by $[q,\chi_1,\chi_2,e_{\rm ref},\zeta_{\rm ref}]=[2,0.4,-0.2,0.15,\pi/2]$. For comparison, we also include the $(2,2)$ (\textit{second panel}) mode spectrogram from Figure~\ref{fig:spectrogram}. For comparison, we show the circular time-frequency tracks in red. Details are in Section~\ref{sec:time-frequency}.}
\label{fig:spectrogram_hm}
\end{figure}

\begin{figure}
\includegraphics[width=\columnwidth]{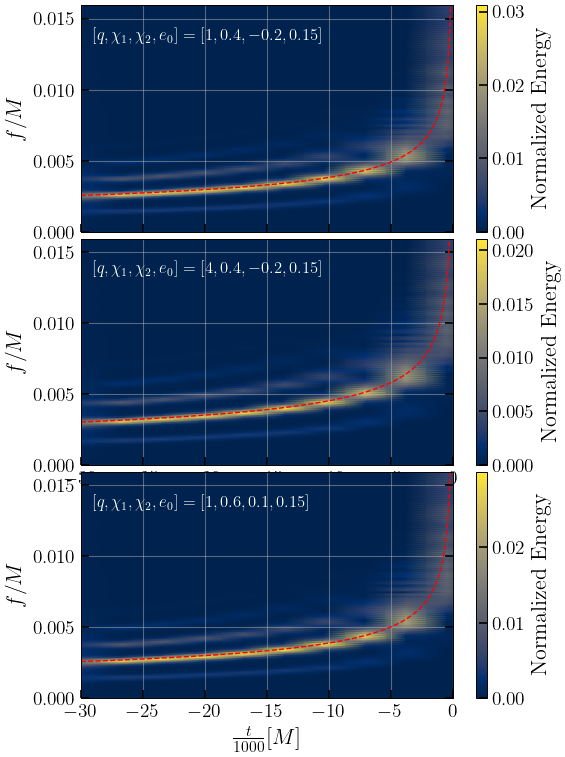}
\caption{We show time-frequency spectrogram obtained from the short-time Fourier transform (STFT) for the $(2,2)$ mode for three representative spinning eccentric binary characterized by $[q,\chi_1,\chi_2,e_{\rm ref}]=[1,0.4,-0.2,0.15]$, $[4,0.4,-0.2,0.15]$ and $[1,0.6,0.1,0.15]$. We fix the relativisitc anomaly to be $\zeta_{\rm ref}=\pi/2$. For comparison, we show the circular time-frequency tracks in red. Details are in Section~\ref{sec:time-frequency}.}
\label{fig:spectrogram_q1q4}
\end{figure}

\section{Notation}
\label{sec:notation}
The full complex waveform $h(t)$ consists of $-2$ spin-weighted spherical harmonic modes $h_{\ell m}$ indexed by $(\ell,m)$~\cite{Maggiore:2007ulw,Maggiore:2018sht}:
\begin{align}
h(t,\theta,\phi;\boldsymbol{\lambda}) &= \sum_{\ell=2}^\infty \sum_{m=-\ell}^{\ell} h_{\ell m}(t;\boldsymbol{\lambda})\,{}_{-2}Y_{\ell m}(\theta,\phi).
\label{hmodes}
\end{align}
Here, $t$ denotes time, $\theta$ and $\phi$ are the usual spherical angles on the sky, and $\boldsymbol{\lambda}$ is the set of intrinsic parameters:
$\boldsymbol{\lambda} = \{q,\chi_1,\chi_2,e_{\rm ref},l_{\rm ref}\}$,
where $q = m_1/m_2$ is the mass ratio (with $m_1 \ge m_2$), and $\chi_1$, $\chi_2$ are the dimensionless spin components along the orbital angular momentum for non-precessing binaries. Instead of a standardized definition of eccentricity, we adopt waveform model’s internal definition for $e_{\rm ref}$ and reference mean anomaly $l_{\rm ref}$. Instead of $l_{\rm ref}$, one can use relativistic anomaly $\zeta(t)$ (e.g. in \texttt{SEOBNRv5EHM} model). We use geometric units ($G = c = 1$), measure time in units of the total mass $M = m_1 + m_2$, align the waveform such that the peak amplitude of the $(2,2)$ mode occurs at $t = 0$, and set the phase to zero at the start of the waveform.
In most binaries, the dominant radiation arises from the $(\ell,m)=(2,2)$ mode—commonly called the quadrupolar mode—while the remaining modes are referred to as higher-order modes. Each mode has a real-valued amplitude $A_{\ell m}(t)$ and phase $\phi_{\ell m}(t)$, so that
$h_{\ell m}(t;\boldsymbol{\lambda}) = A_{\ell m}(t)\,e^{i\phi_{\ell m}(t)}$.
The instantaneous angular frequency of each mode is
$\omega_{\ell m}(t;\boldsymbol{\lambda}) = {d\phi_{\ell m}(t)}/{dt}$,
and the corresponding GW frequency is $f_{\ell m}(t) = {\omega_{\ell m}(t)}/{\pi}$.

For eccentric binaries, each spherical harmonic mode can be further decomposed into smooth harmonics that depend solely on eccentricity~\cite{Yunes:2009yz,Patterson:2024vbo,Islam:2025rjl}:
\begin{align}
h_{\ell m}(t;\boldsymbol{\lambda}) &= \sum_{j=j_{\rm min}}^{\infty} h_{\ell m,j}(t;\boldsymbol{\lambda}),
\label{eq:eccentric_decomposition}
\end{align}
where $j$ indexes the eccentric harmonics. In the quasi-circular limit, only one eccentric harmonic survives. Once the eccentric harmonics are obtained, their amplitude, phase, instantaneous angular frequency, and GW frequency (for the $(\ell,m)$ mode) are given by $A_{\ell m,j}(t) = |h_{\ell m,j}|$, 
$\phi_{\ell m,j}(t) = {\rm Arg}[h_{\ell m,j}]$, $\omega_{\ell m,j}(t) = {d\phi_{\ell m,j}}/{dt}$ and $f_{\ell m,j}(t) = {\omega_{\ell m,j}}/{\pi}$. 
In PAPER-I, we have showed that for the quadrupolar mode, $j=2$ is the dominant eccentric harmonic. Moreover, the phases and frequencies of the eccentric harmonics in the $(2,2)$ mode comprise two distinct components: one associated with the orbital motion and the other arising solely from eccentricity. We can express these as
\begin{align}
\phi_{22,j}(t) &\approx j\,\phi_{\lambda}(t) + \phi_{\rm ecc}(t),\\
\omega_{22,j}(t) &\approx j\,\omega_{\lambda}(t) + \omega_{\rm ecc}(t),\\
f_{22,j}(t) &\approx j\,f_{\lambda}(t) + f_{\rm ecc}(t),
\label{eq:ecc_phase_22}
\end{align}
where $\phi_{\lambda}(t)$ (with $\omega_{\lambda}(t)={d\phi_{\lambda}}/{dt}$ and $f_{\lambda}(t)={\omega_{\lambda}}/{\pi}$) denotes the secular orbital phase in the circular limit, scaling linearly with $j$, while $\phi_{\rm ecc}(t)$, $\omega_{\rm ecc}(t)$, and $f_{\rm ecc}(t)$ depend only on eccentricity and are independent of $j$. For the $j=3$ harmonic, an additional $\pi$ phase offset is also observed.

\section{Time–Frequency Spectrograms}
\label{sec:time-frequency}
First, we examine the structure of eccentric harmonics across various eccentric, spinning binaries and spherical-harmonic modes using time–frequency spectrograms obtained via the short-time Fourier transform (STFT). We compute the STFT using the \texttt{scipy.signal.stft} module. For demonstration, all waveforms in this paper are generated with durations of approximately $30{,}000\,M$.

\begin{figure}
\includegraphics[width=\columnwidth]{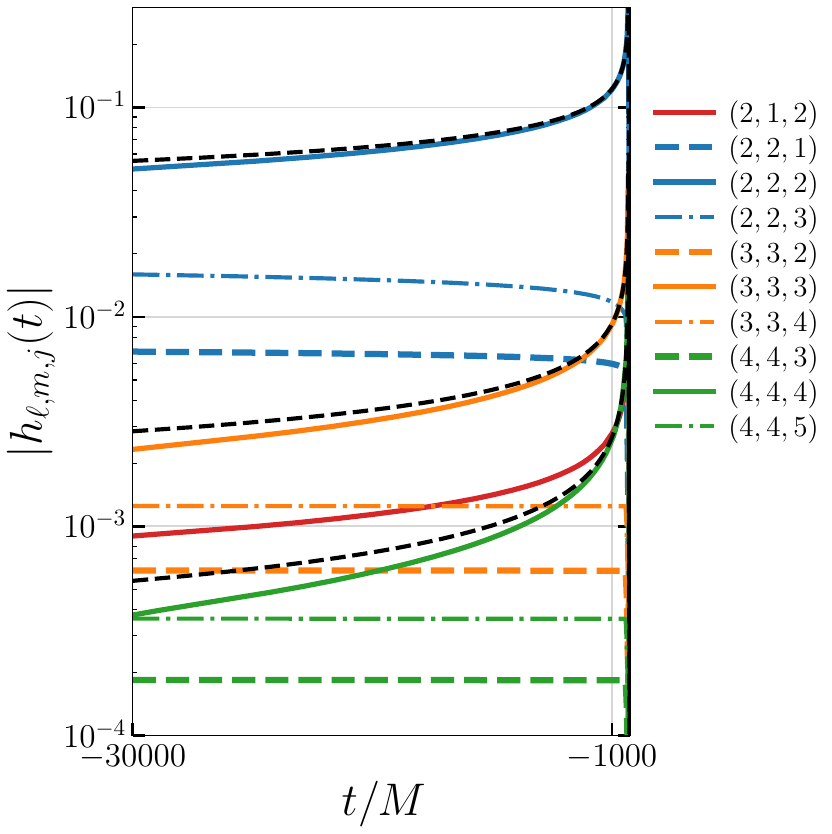}
\caption{We show the amplitudes of several eccentric harmonics $(\ell, |m|, j)$ extracted from the $(\ell, |m|) = (2,2)$, $(3,3)$ and $(4,4)$ modes for the binary characterized by $[q,\chi_1,\chi_2,e_{\rm ref},\zeta_{\rm ref}]=[2,0.4,-0.2,0.15,\pi/2]$. For comparison, the corresponding circular amplitudes for the $(\ell, |m|) = (2,2)$, $(3,3)$ and $(4,4)$ modes are plotted as black dashed lines. Details are provided in Section~\ref{sec:extraction}.}
\label{fig:amp_hierarchy}
\end{figure}

\begin{figure}
\includegraphics[width=\columnwidth]{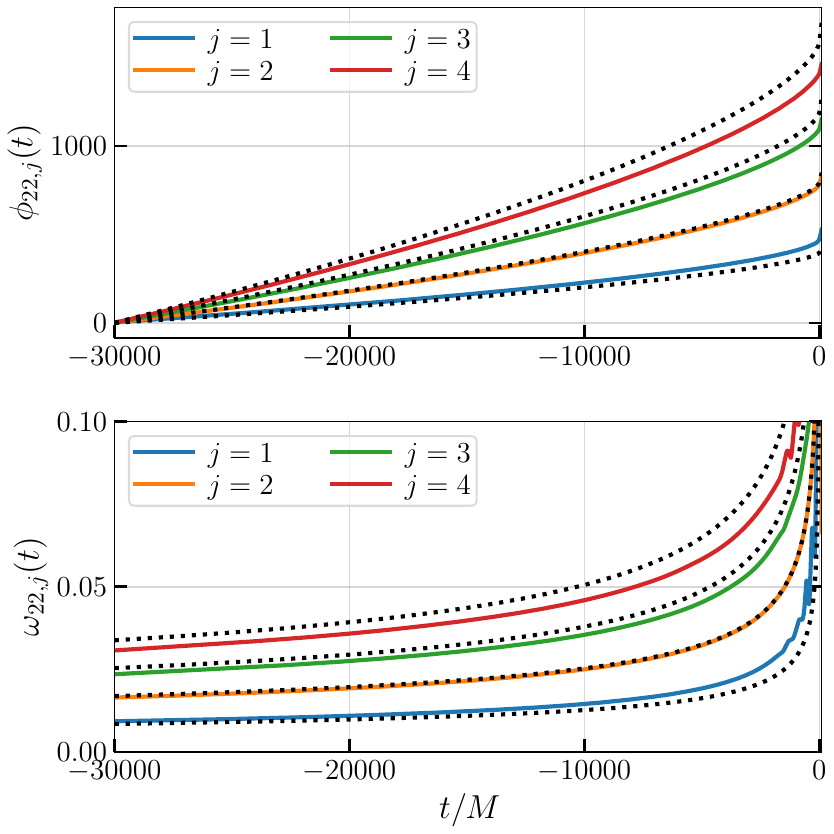}
\caption{In our model, the phases and frequencies of the eccentric harmonics are expressed as a sum of secular orbital phase and an eccentric component (see Eq.~\ref{eq:ecc_phase_22}). We show the secular component of the phases (upper panel) and instantaneous frequencies (lower panel) of the eccentric harmonics with $j = \{1,2,3,4\}$ extracted from the $(2,2)$ mode for the binary characterized by $[q,\chi_1,\chi_2,e_{\rm ref},\zeta_{\rm ref}]=[2,0.4,-0.2,0.15,\pi/2]$. For comparison, the corresponding multiples of the circular orbital phase and frequency are plotted as dotted lines (which do not agree well with the solid lines due to the effect of periastron advance). Further details are provided in Section~\ref{sec:extraction}.}
\label{fig:ecc_harm_hierarchy}
\end{figure}

We begin with the $(2,2)$ mode of the gravitational waveform. In Fig.~\ref{fig:spectrogram}, we show the real part of the $(2,2)$ mode generated by two spinning eccentric binaries characterized by $[q,\chi_1,\chi_2,e_{\rm ref},\zeta_{\rm ref}]=[2,0.4,-0.2,0.8,\pi/2]$ (maroon line) and $[2,0.4,-0.2,0.15,\pi/2]$ (blue line) using the \texttt{SEOBNRv5EHM} model. The reference time is chosen as $t\approx-30000\,M$, while $t=0$ denotes the merger. We observe that extreme eccentricities ($e_{\rm ref}=0.8$) yield pronounced modulations in the waveform, whereas moderate eccentricities produce much smaller modulations. For $e_{\rm ref}=0.8$, the time–frequency spectrogram reveals several overlapping structures. However, for $e_{\rm ref}=0.15$, these time–frequency features become much clearer, exhibiting clean tracks corresponding to individual eccentric harmonics. These structures closely resemble those previously shown in Refs.~\cite{Islam:2025rjl,Patterson:2024vbo}. Near merger ($t=-3000\,M$), these distinct tracks overlap. We find that for moderate eccentricities, this qualitative behavior persists.

We then inspect the time–frequency spectrograms of the higher-order modes. In Fig.~\ref{fig:spectrogram_hm}, we show spectrograms for four modes—$(2,1)$ (first panel), $(2,2)$ (second panel), $(3,3)$ (third panel), and $(4,4)$ (fourth panel)—for the same binary. For comparison, the circular time–frequency tracks are shown in red. There are a couple of points to note. First, the normalized energies in the higher-order modes are smaller than in the $(2,2)$ mode, as expected. Each mode exhibits several time–frequency tracks corresponding to different eccentric harmonics. Notably, as we consider higher-order modes, these harmonics start to overlap much earlier than in the $(2,2)$ mode. The only exception is the $(2,1)$ mode, in which the eccentric harmonics overlap later than in the $(2,2)$ mode. This indicates that eccentric harmonics decay more rapidly in higher-order modes and that their decay depends on $\ell$ and $m$.

Next, we study how the time–frequency spectrogram changes with mass ratio and spin while keeping the eccentricity fixed ($e_{\rm ref}=0.15$). In Figure~\ref{fig:spectrogram_q1q4}, we show the spectrograms of the $(2,2)$ mode for three representative spinning eccentric binaries characterized by $[q,\chi_1,\chi_2,e_{\rm ref}]=[1,0.4,-0.2,0.15]$, $[4,0.4,-0.2,0.15]$, and $[1,0.6,0.1,0.15]$. We fix the relativistic anomaly at $\zeta_{\rm ref}=\pi/2$. For comparison, the circular time–frequency tracks are shown in red. The first two spectrograms illustrate the effect of mass ratio, while the first and third illustrate the effect of spin. We observe that, despite different mass ratio and spin configurations, all spectrograms appear similar. This further demonstrates that these harmonics depend primarily on eccentricity and only weakly on mass ratio or spin.

\begin{figure}
\includegraphics[width=\columnwidth]{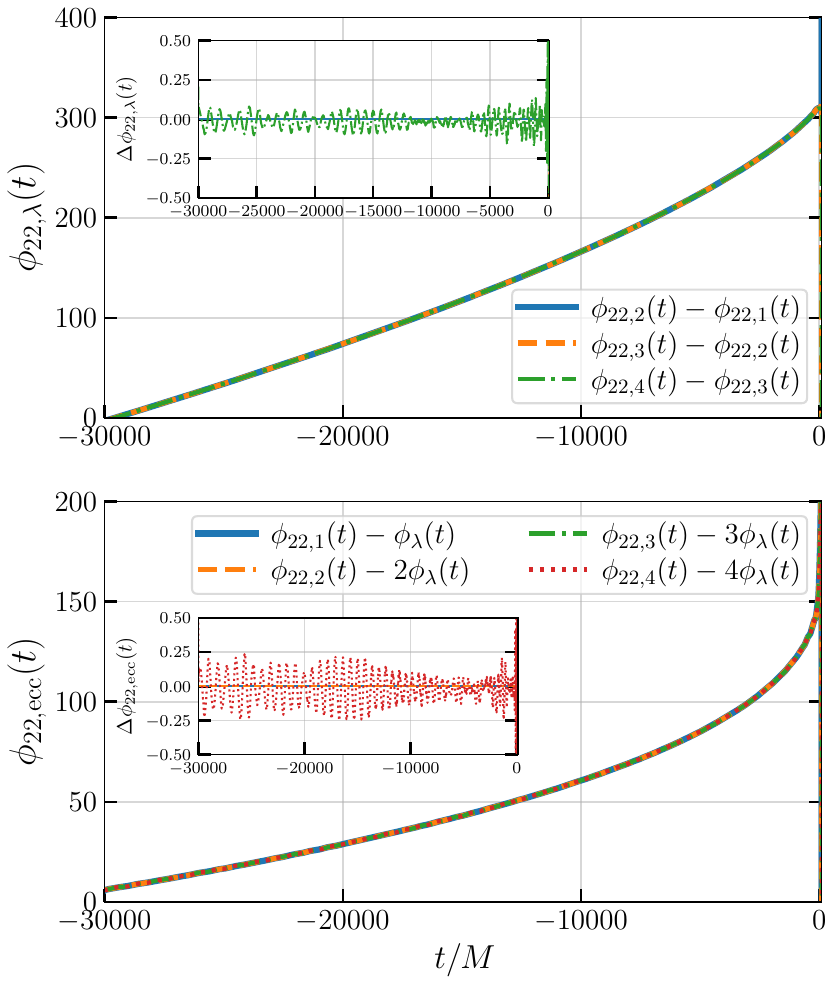}
\caption{Test of robustness of our phase model in Eq.~\ref{eq:ecc_phase_22}. We show the phase relation between different eccentric harmonics for the signal shown in Fig.~\ref{fig:spectrogram_hm}. The upper panel shows $\phi_{\lambda}(t)$ (which is the leading order phase tied to the orbital motion) while the lower panel presents $\phi_{\rm ecc}(t)$ (which is related to periastron advance) computed from the quadrupolar mode. We show the residuals in the inset. Details are provided in Section~\ref{sec:extraction}. 
}
\label{fig:ecc_harm_phase_relation}
\end{figure}

\section{Extraction \& Phenomenology}
\label{sec:extraction}

We now extract the eccentric harmonics from the fiducial waveform generated using the \texttt{SEOBNRv5EHM} model for $[q,\chi_1,\chi_2,e_{\rm ref},\zeta_{\rm ref}]=[2,0.4,-0.2,0.15,\pi/2]$. The spectrograms of several representative modes are shown in Figure~\ref{fig:spectrogram_hm}. We obtain the eccentric harmonics using the \texttt{gwMiner} package’s \texttt{svd\_method} described in PAPER-I. Although the package also offers a heterodyning-based method, it results in loss of signal at the beginning and end due to the Gibbs phenomenon; we therefore do not employ it here. PAPER-I provides a detailed description of the \texttt{svd\_method}, but we summarize it briefly. For the binary with $[q,\chi_1,\chi_2,e_{\rm ref},\zeta_{\rm ref}]=[2,0.4,-0.2,0.15,\pi/2]$, we generate an ensemble of 50 waveforms sharing $e_{\rm ref}=0.15$ but with $\zeta_{\rm ref}$ uniformly sampled in $[0,2\pi]$. 
We first solve an optimization problem to determine the initial orbital frequency supplied to \texttt{SEOBNRv5EHM} such that, at the reference time $t=-30{,}000\,M$, the parameters satisfy $e_{\rm ref}$ and $\zeta_{\rm ref}=\pi/2$. We then use this same initial frequency to generate all 50 waveforms with different $\zeta_{\rm ref}$. Note that this reference choice is not unique; a slightly different reference point could be adopted.

We then interpolate these 50 waveforms onto a common time grid, align their starting phases to zero, and apply SVD; the resulting bases correspond to eccentric harmonics. We then compute their frequencies to assign each harmonic index. Finally, we smooth the harmonics using the PN-inspired amplitude fit presented in PAPER-I. For higher-order modes, since the eccentric harmonics begin to overlap early in time, we use only the first $5000\,M$ of the raw harmonics when smoothing their amplitudes. This prevents power spillover from dominant to subdominant harmonics—due to SVD imperfections---from violating the expectation that harmonic amplitudes decay monotonically as eccentricity decreases. Additionally, in PAPER-I, using the \texttt{TEOBResumS} model, the waveform interface did not expose the mean (relativistic) anomaly directly; we therefore employed a PN-based eccentricity–frequency evolution track (cf.\ Sec.~II.A.2; Fig.~2) to generate waveforms with varying mean anomaly at fixed eccentricity. In contrast, \texttt{SEOBNRv5EHM} provides direct access to the mean (relativistic) anomaly, allowing us to generate the waveform ensemble without an eccentricity–frequency track.

\begin{figure}
\includegraphics[width=\columnwidth]{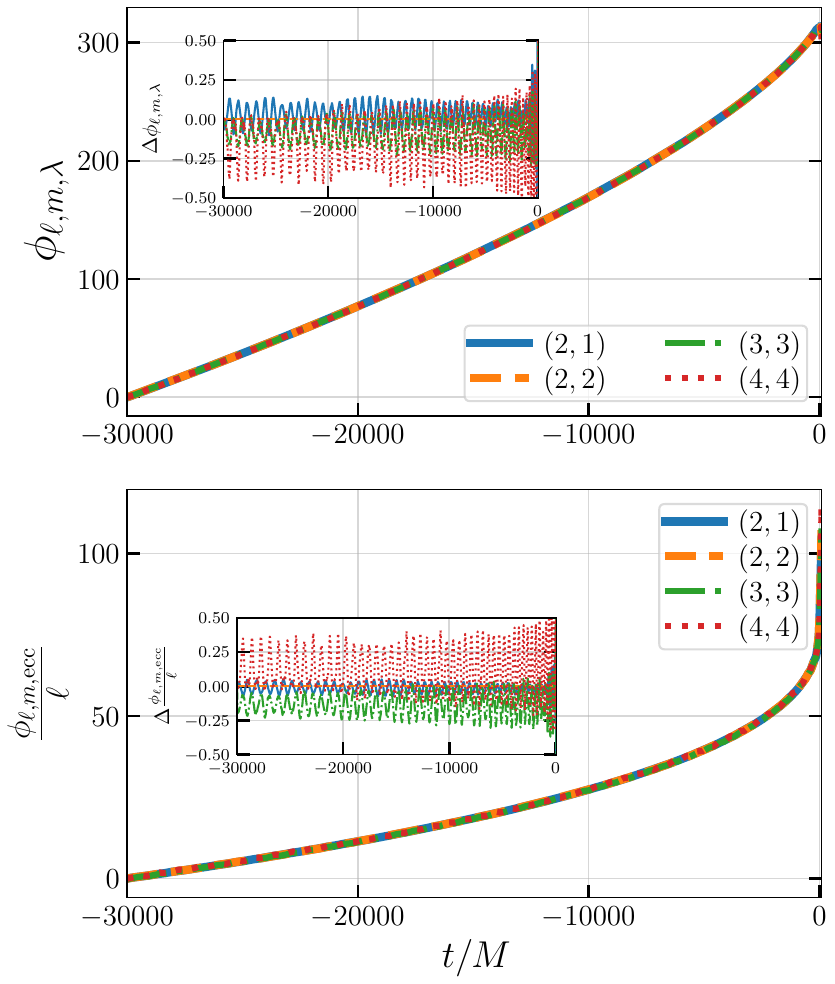}
\caption{Same as Fig.~\ref{fig:ecc_harm_phase_relation} but for harmonics other than the quadrupole. We show the phase relation given by Eq.~\ref{eq:ecc_phase_hm} between different eccentric harmonics for the signal shown in Fig.~\ref{fig:spectrogram_hm} for the binary characterized by $[q,\chi_1,\chi_2,e_{\rm ref},\zeta_{\rm ref}]=[2,0.4,-0.2,0.15,\pi/2]$. The upper panel shows $\phi_{\lambda}(t)$ while the lower panel presents $\phi_{\rm ecc}(t)$ computed from different extracted modes. We show the residuals in the inset. Details are provided in Section~\ref{sec:extraction}. 
}
\label{fig:ecc_harm_phase_relation_hm}
\end{figure}

Note that SVD returns the normalized eccentric-harmonic basis $e_{\ell m,j}(t)$ (cf.\ PAPER-I). Any eccentric harmonic mode for a given $(q,\chi_1,\chi_2,e_{\rm ref})$ can be written as
\begin{equation}
h_{\ell m,j}(t) = C_{j}(\zeta_{\rm ref})\,e_{\ell m,j}(t; q,\chi_1,\chi_2,e_{\rm ref}),
\label{eq:full_ecc_harmonic_decomposition}
\end{equation}
where $C_{j}(\zeta_{\rm ref})$ are complex SVD coefficients that depend only on the (relativistic) mean anomaly $\zeta_{\rm ref}$ for the ensemble (since $q$, $\chi_1$, $\chi_2$, and $e_{\rm ref}$ are fixed), and $e_{\ell m,j}(t)$ are the normalized eccentric-harmonic basis functions. We obtain the coefficients $C_{j}(\zeta_{\rm ref})$ by projecting each waveform $h_{\ell m}(t; q,\chi_1,\chi_2,e_{\rm ref},\zeta_{\rm ref})$ onto the normalized basis $\{e_{\ell m,j}\}$.

We find that the dominant eccentric harmonic index depends on the spherical-harmonic mode: for each $(\ell,m)$ mode, the dominant index is $j = m$. Furthermore, for the $(2,2)$ mode, the first eccentric harmonic is indexed by $j = 1$ and we choose to work with the first four harmonics. For all modes, the eccentric harmonic (that has reasonable signal contribution) indices begin at $j = m - 1$.
We then show the amplitudes of several eccentric harmonics extracted from the $(2,2)$, $(3,3)$ and $(4,4)$ modes in Figure~\ref{fig:amp_hierarchy}. For comparison, the corresponding circular amplitudes are plotted as dotted lines. We find that the $(\ell,m,j)=(2,2,1)$ and $(\ell,m,j)=(2,2,3)$ are dominant over the $(\ell,m,j)=(3,3,3)$ eccentric harmonic.
Below we provide the overall amplitude dependence on the eccentricity for the sub-dominant eccentric harmonics for the $(2,2)$ mode:
\begin{align}
    A_{2,2,1} \sim 0.0022 e^{0.1}(t),\\ 
    A_{2,2,3} \sim 0.0031 e^{0.24}(t),\\
    A_{2,2,4} \sim 0.0247 e^{1.20}(t),
\end{align}
and for the other modes:
\begin{align}
    A_{2,1,5} \sim 0.011 e^{0.82}(t),\\
    A_{3,3,5} \sim 0.011 e^{0.82}(t),\\
    A_{4,4,6} \sim 0.004 e^{0.37}(t).
\end{align}
Some other modes exhibit amplitudes $A_{\ell,m,j}\sim e(t)$ at the early inspiral and we carry that forward at later times.

Next, we observe that the phases and instantaneous frequencies of the eccentric harmonics for the quadrupolar mode in spinning eccentric binaries obey the same relations [Eq.~(\ref{eq:ecc_phase_22})] as those derived in PAPER-I for the non-spinning eccentric limit using the \texttt{TEOBResumS} model. We demonstrate this result in Figure~\ref{fig:ecc_harm_hierarchy}, which presents the phases and frequencies of the first four eccentric harmonics alongside their circular counterparts. Figure~\ref{fig:ecc_harm_phase_relation} then compares $\phi_{\lambda}(t)$ and $\phi_{\rm ecc}(t)$ obtained from different harmonic combinations and shows their mutual consistency. We also compute the residuals between them and find them to be only $\sim0.25\,$rad.

Next, we inspect this phase relation across different spherical-harmonic modes and find that each mode individually obeys the same phase–frequency relation. Hence, one can write
\begin{align}
\phi_{\ell m,j}(t) &\approx j\,\phi_{\ell m,\lambda}(t) + \phi_{\ell m,\rm ecc}(t),\\
\omega_{\ell m,j}(t) &\approx j\,\omega_{\ell m,\lambda}(t) + \omega_{\ell m,\rm ecc}(t),\\
f_{\ell m,j}(t) &\approx j\,f_{\ell m,\lambda}(t) + f_{\ell m,\rm ecc}(t),
\label{eq:ecc_phase_hm}
\end{align}
where the secular phase component $\phi_{\ell m,\lambda}(t)$ is the same for all modes, while the eccentric component $\phi_{\ell m,\rm ecc}(t)$ varies by mode. However, when scaled by $\ell/2$, these eccentric components collapse onto a single curve. The same scaling holds for $\omega_{\ell m,\rm ecc}(t)$ and $f_{\ell m,\rm ecc}(t)$, suggesting that $\phi_{\ell m,\lambda}(t)$ corresponds to a common orbital phase. We can therefore drop the $(\ell,m)$ subscript and write
\begin{align}
\phi_{\ell m,j}(t) &\approx j\,\phi_{\lambda}(t) + \frac{\ell}{2}\,\phi_{\rm ecc}(t),\\
\omega_{\ell m,j}(t) &\approx j\,\omega_{\lambda}(t) + \frac{\ell}{2}\,\omega_{\rm ecc}(t),\\
f_{\ell m,j}(t) &\approx j\,f_{\lambda}(t) + \frac{\ell}{2}\,f_{\rm ecc}(t),
\label{eq:ecc_phase_hm_simplified}
\end{align}
with $\phi_{\rm ecc}(t)\equiv\phi_{22,\rm ecc}(t)$. Interestingly, Refs.~\cite{Islam:2024rhm,Islam:2024bza} observed a similar $\ell/2$ scaling of universal eccentric modulations across modes. We demonstrate these relations by plotting $\phi_{\lambda}(t)$ and $\phi_{\rm ecc}(t)$ obtained from different spherical-harmonic modes in Figure~\ref{fig:ecc_harm_phase_relation_hm}. We also compute the residuals between them and find them to be only $\sim0.5\,$rad.

\begin{figure}
\includegraphics[width=\columnwidth]{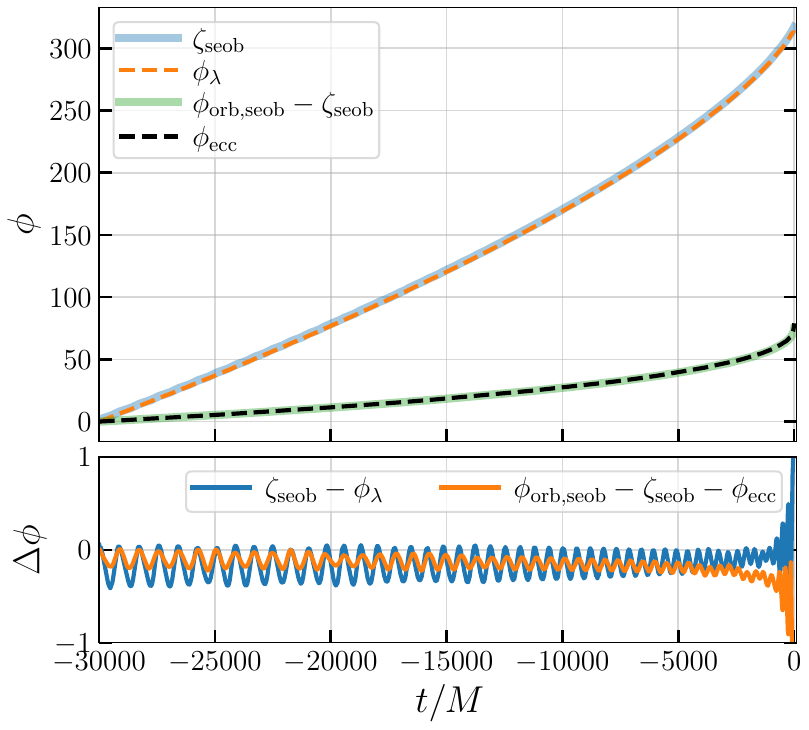}
\caption{We show the relationship between $\phi_{\lambda}(t)$ and $\phi_{\rm ecc}(t)$ (obtained from \texttt{gwMiner}) with the orbital angle $\phi_{\rm orb, seob}(t)$ and relativistic anomaly $\zeta(t)$ obtained from the \texttt{SEOBNRv5EHM} dynamics for the binary characterized by $[q,\chi_1,\chi_2,e_{\rm ref},\zeta_{\rm ref}]=[2,0.4,-0.2,0.15,\pi/2]$. The upper panel shows the phases, while the lower panel shows the residual. Details are provided in Section~\ref{sec:extraction}.
}
\label{fig:seob_mean_ano}
\end{figure}

Next, we connect the data-driven phases $\phi_{\lambda}(t)$ and $\phi_{\rm ecc}(t)$ extracted from the \texttt{SEOBNRv5EHM} waveforms to PN-based dynamics. We obtain the necessary dynamics directly from the \texttt{SEOBNRv5EHM} model \cite{Gamboa:2024imd}. PN literature suggests that, as found here and in PAPER-I, the phase of any eccentric spherical-harmonic mode comprises two primary components: a secular orbital phase and an eccentric precession phase. The secular phase is given by the mean anomaly $l(t)$ or $\zeta(t)$, while the eccentric precession phase combines $\zeta(t)$ and the periastron advance $k(t)$. The \texttt{SEOBNRv5EHM} dynamics provide both $\zeta(t)$ and the total orbital phase $\phi_{\rm orb, seob}$. We thus compute
\[
\phi_{\rm prec,seob}(t) \equiv \phi_{\rm orb, seob}(t) - \zeta_{\rm seob}(t),
\]
which corresponds to the precession phase in the model. Although both $\phi_{\rm orb, seob}$ and $\zeta(t)$ are orbit-averaged quantities, they retain PN-related oscillations due to eccentricity; a true orbit-average would remove such oscillations. Figure~\ref{fig:seob_mean_ano} displays $\phi_{\lambda}(t)$ and $\phi_{\rm ecc}(t)$ alongside $\phi_{\rm orb, seob}(t)$ and $\zeta(t)$. We find
\[
\phi_{\lambda}(t) \approx \zeta_{\rm seob}(t), 
\qquad
\phi_{\rm ecc}(t) \approx \phi_{\rm orb, seob}(t) - \zeta_{\rm seob}(t),
\]
with residuals below $1\,$rad. We note that similar attempts were made to match $\phi_{\lambda}(t)$ and $\phi_{\rm ecc}(t)$ to the 3PN calculations of the mean anomaly $l(t)$ and precession advance $k(t)$ from Ref.~\cite{Moore:2016qxz}. While we found $\phi_{\lambda}(t)\approx l(t)$, noticeable discrepancies remained between $\phi_{\rm ecc}(t)$ and the 3PN prediction $k(t)\,l(t)$. In contrast, \texttt{SEOBNRv5EHM} incorporates 3PN effects in a resummed form, yielding smoother behavior around merger. Moreover, it provides a consistent test, since both waveform and dynamics derive from the same model.

\begin{figure}
\includegraphics[width=\columnwidth]{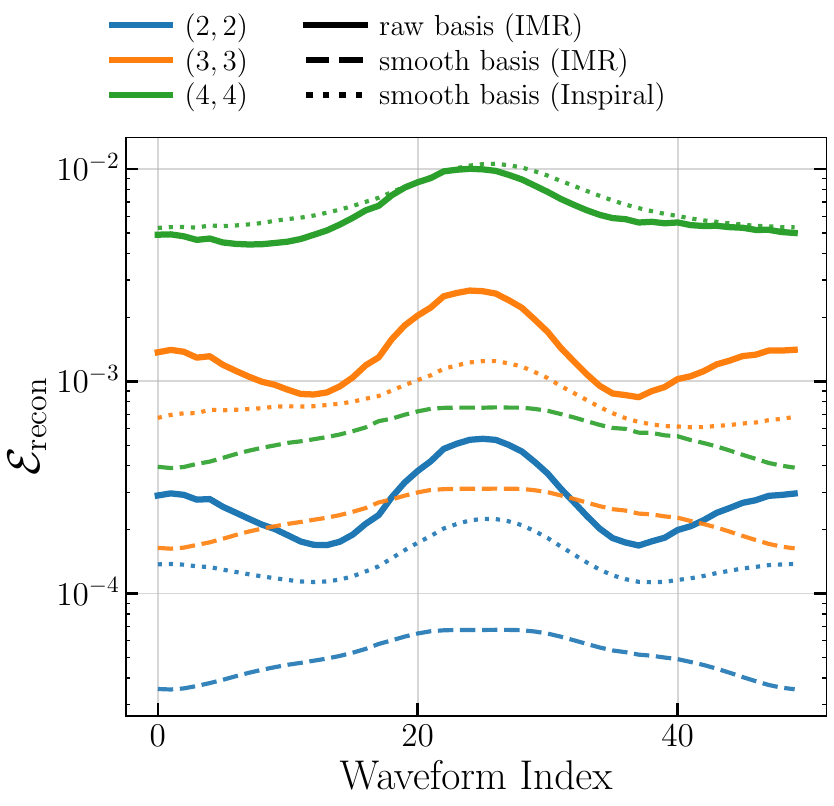}
\caption{We show the reconstruction errors for three different representative modes, $(2,2)$, $(3,3)$ and $(4,4)$, for all 50 waveforms used in the SVD-based method to find the eccentric harmonics. These waveforms are characterized by $[q,\chi_1,\chi_2,e_{\rm ref}]=[2,0.4,-0.2,0.15]$ (with different $\zeta_{\rm ref}$ values).  We use first four eccentric harmonic bases (both the raw and smooth ones). Furthermore, we show the reconstruction error if we only consider the inspiral part of the waveform (i.e. $t\leq-100M$). Details are provided in Section~\ref{sec:extraction}.
}
\label{fig:recon_error}
\end{figure}

Finally, we compute the reconstruction errors for three representative modes—$(2,2)$, $(3,3)$, and $(4,4)$—for all 50 waveforms used in the eccentric-harmonic extraction. We employ only the first four eccentric harmonics, both raw (directly from SVD) and smoothed (using the PN-inspired amplitude fits). To assess errors in the merger regime, we also compute the reconstruction error restricted to the inspiral portion ($t \le -100\,M$). We use the relative $L_2$-norm error as our metric~\cite{blackman2017numerical,Islam:2021mha}. We find that the reconstruction error for the $(2,2)$ mode is $\sim10^{-4}$, indicating excellent agreement; smoothing the SVD bases moderately increases this error. For the $(3,3)$ and $(4,4)$ modes, the errors rise to $\sim10^{-3}$ and $\sim10^{-2}$, respectively, suggesting that extracting eccentric harmonics from higher-order modes is more challenging. This trend aligns with our observation in Section~\ref{sec:time-frequency}, where significant overlap between eccentric-harmonic time–frequency tracks occurs earlier in higher-order modes than in the $(2,2)$ mode. Furthermore, for the $(2,2)$ mode, we find that the reconstruction errors for spinning eccentric binaries are comparable to those obtained for non-spinning eccentric binaries in PAPER-I (cf. Fig.~7 therein).

Although most results in this Section are shown for $[q,\chi_1,\chi_2,e_{\rm ref},\zeta_{\rm ref}]=[2,0.4,-0.2,0.15,\pi/2]$, we find that these observations hold for any spinning binary with moderate eccentricity ($e_{\rm ref}\leq0.4$).

\begin{figure}
\includegraphics[width=\columnwidth]{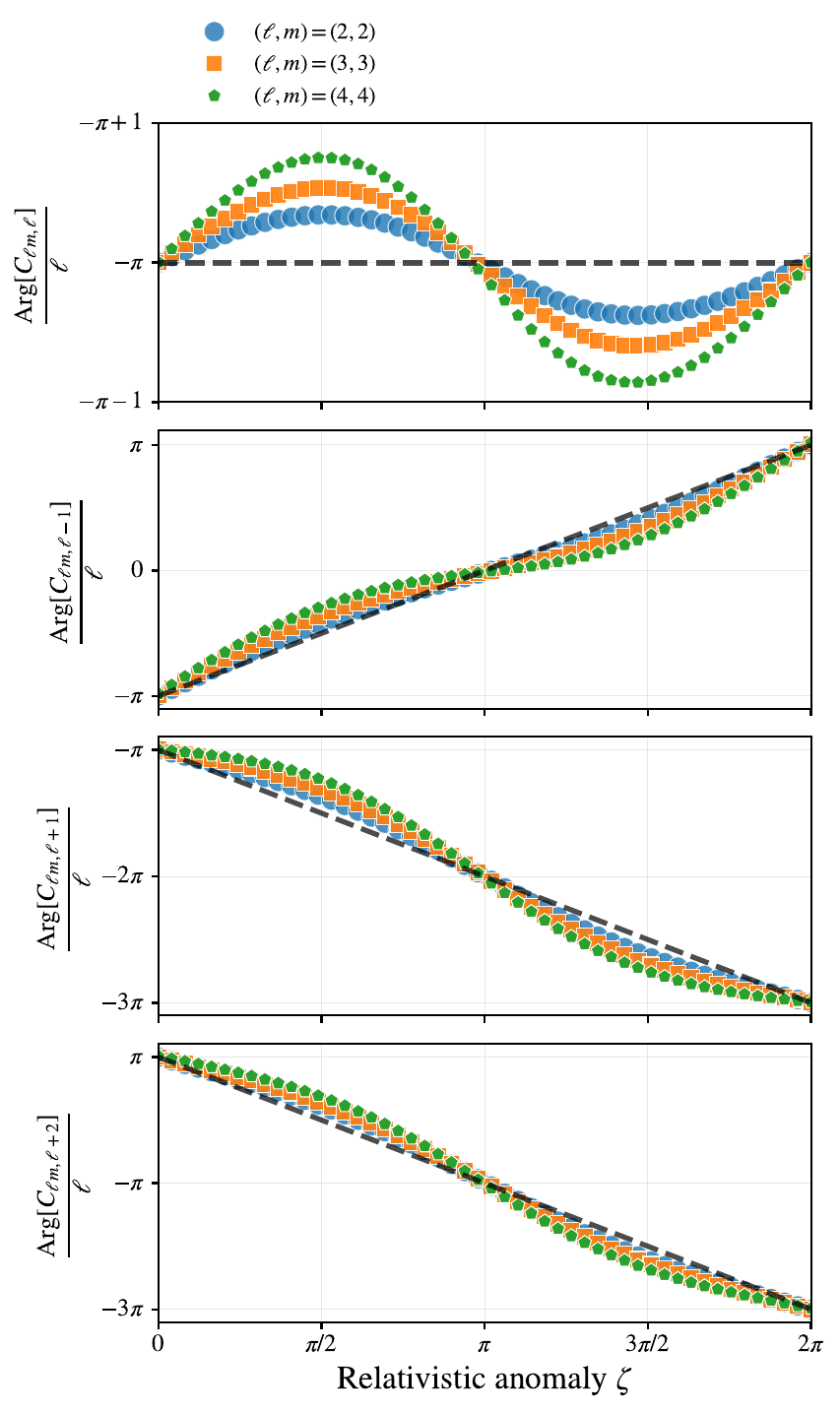}
\caption{In our model for the full eccentric waveforms from Eq.~\ref{eq:full_ecc_harmonic_decomposition}, the SVD coefficient $C_{j}(\zeta_{\rm ref})$ captures the entire dependence on the relativistic mean anomaly $\zeta$. In this plot, we present the phases (arguments) of the SVD coefficients $C_{\ell m j}$ for the first four eccentric harmonics, $j \in \{\ell-1,\ell,\ell+1,\ell+2\}$, obtained from the $(\ell,m)$ spherical-harmonic mode, plotted as differently coloured markers, with leading-order fits shown as dashed black lines. These values correspond to a BBH characterized by $[q,\chi_1,\chi_2,e_{\rm ref}]=[2,0.4,-0.2,0.15]$ and are qualitatively similar across the BBHs considered in this work. Details are in Section~\ref{sec:model_mean_anomaly}.}
\label{fig:ph_clmj}
\end{figure}

\section{Investigating the effect of mean anomaly}
\label{sec:model_mean_anomaly}
We now investigate the behaviour of the complex SVD coefficients $C_{\ell m j}(\zeta_{\rm ref})$ as we vary the relativistic anomaly $\zeta_{\rm ref}$ for a fixed $(q,\chi_1,\chi_2,e_{\rm ref})$. Specifically, we examine the coefficients $C_{\ell m j}(\zeta_{\rm ref})$ obtained for each eccentric BBH in Section~\ref{sec:extraction}. A similar analysis was carried out in PAPER-I (for non-spinning eccentric binaries) for the quadrupolar mode. Here, we extend that study to higher-order spherical-harmonic modes and to spinning binaries, assessing whether the results of PAPER-I apply across modes within the \texttt{SEOBNRv5EHM} model. Since $C_{\ell m j}(\zeta_{\rm ref})$ are complex-valued, we model the amplitudes and phases separately.

\begin{figure}
\includegraphics[width=\columnwidth]{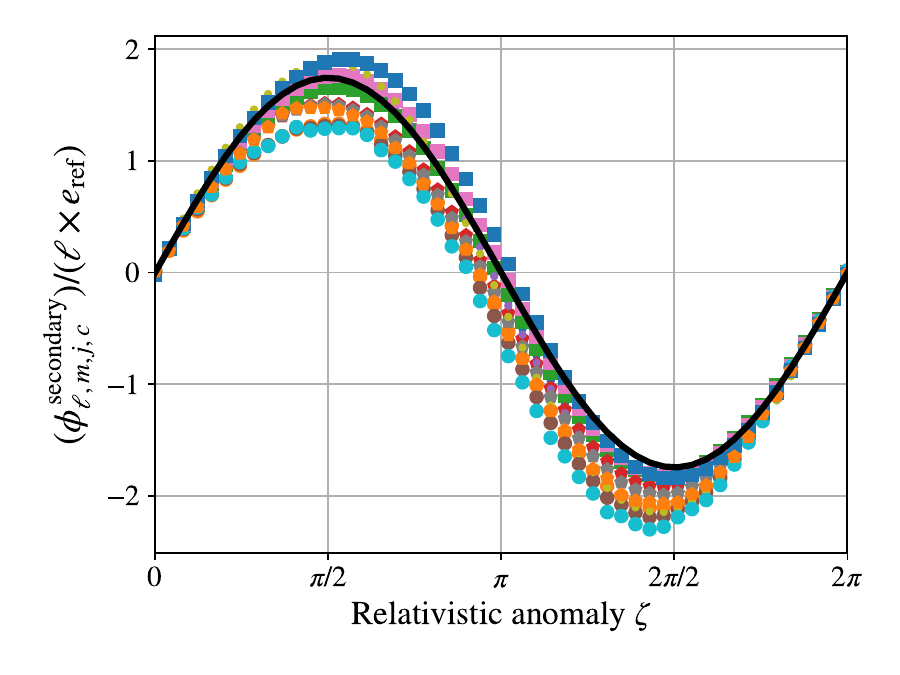}
\caption{Subtracting the leading-order fits shown in Fig.~\ref{fig:ph_clmj}, we obtain the phase of the residual secondary oscillatory phase component (Eq.~\ref{eq:leading_secondary_decomposition}). We show the this component, $\phi_{\ell m j}^{\rm secondary}$, as a function of the relativistic mean anomaly $\zeta_{\rm ref}$. Different markers show values computed from the $(2,2)$, $(3,3)$, and $(4,4)$ modes for four binaries with varying $q$, $\chi_1$, $\chi_2$, and $e_{\rm ref}$. While the amplitude of $\phi_{\ell m j}^{\rm secondary}$ differs with eccentricity and mode indices, the scaled $\phi_{\ell m j}^{\rm secondary}$ exhibits some degree of universality. Our model predictions are shown as a black solid line. Details are in Section~\ref{sec:model_mean_anomaly}.}
\label{fig:ph_clmj_secondary}
\end{figure}

First, for the quadrupolar mode, the argument (phase) of the SVD coefficients $C_{\ell m j}(\zeta_{\rm ref})$ varies similarly to what was found in PAPER-I (with a sign difference). Note, however, that PAPER-I parameterized the dependence with the mean anomaly $l_{\rm ref}$, whereas here we use the relativistic anomaly $\zeta_{\rm ref}$; these are linearly related at leading order~\cite{Gamboa:2024imd}. 
We then examine higher-order spherical-harmonic modes and find that the SVD arguments for the first four harmonics, $j\in\{\ell-1,\ell,\ell+1,\ell+2\}$, exhibit very similar behaviour. 

As in PAPER-I, we find that the dependence of the SVD-coefficient phases on the relativistic anomaly can be decomposed into two components: a leading-order linear term and a secondary oscillatory term,
\begin{equation}
\arg\!\bigl[C_{\ell m j}\bigr](\zeta_{\rm ref})
= \phi_{\ell m j}^{\rm leading}(\zeta_{\rm ref})
+ \phi_{\ell m j}^{\rm secondary}(\zeta_{\rm ref}).
\label{eq:leading_secondary_decomposition}
\end{equation}
The leading behavior is similar across eccentric harmonics obtained from different spherical-harmonic modes:
\begin{align}
\phi_{\ell,m,j=\ell}^{\rm leading}(\zeta_{\rm ref})     &= -\pi,\\
\phi_{\ell,m,j=\ell-1}^{\rm leading}(\zeta_{\rm ref}) &= \,\zeta_{\rm ref}-\pi,\\
\phi_{\ell,m,j=\ell+1}^{\rm leading}(\zeta_{\rm ref}) &= -\zeta_{\rm ref}-\pi,\\
\phi_{\ell,m,j=\ell+2}^{\rm leading}(\zeta_{\rm ref}) &= -2\,\zeta_{\rm ref}-\pi.
\end{align}
In Fig.~\ref{fig:ph_clmj}, we plot the phases (arguments) of $C_{\ell m j}$ for rhe first four important eccentric harmonics indexed by $j\in\{\ell-1,\ell,\ell+1,\ell+2\}$ as colored markers, with the leading-order fits shown as black dashed lines, for a binary corresponding to $[q,\chi_1,\chi_2,e_{\rm ref}]=[2,0.4,-0.2,0.15]$. We confirm that this behavior is same in other binaries too. 

\begin{figure}
\includegraphics[width=\columnwidth]{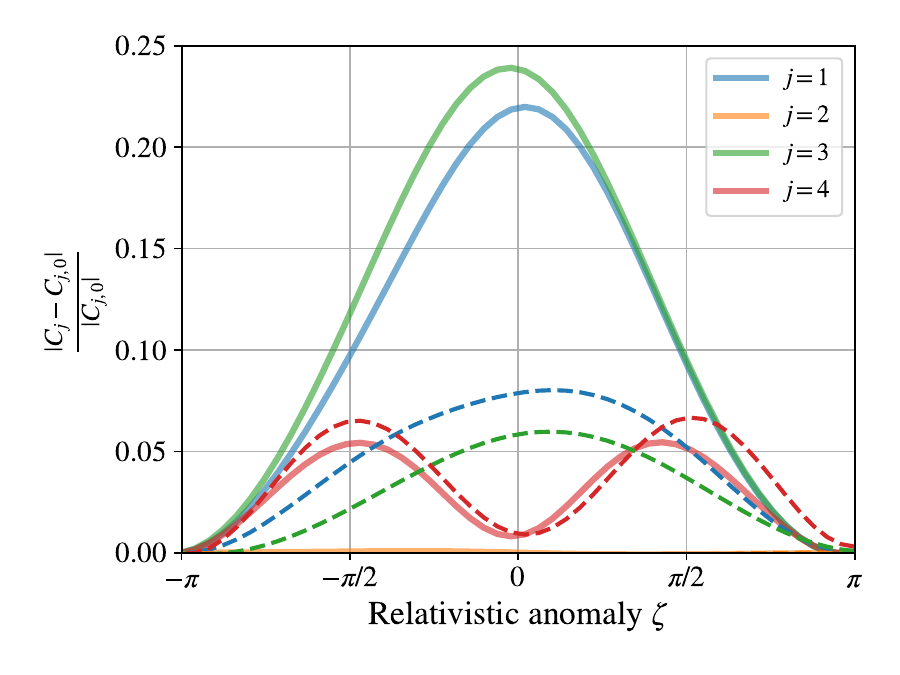}
\caption{We show the relative SVD amplitudes of the eccentric harmonics extracted from the $(2,2)$ mode as functions of the relativistic anomaly for a binary characterized by $[q,\chi_1,\chi_2,e_{\rm ref}]=[2,0,0,0.1]$. Details are in Section~\ref{sec:model_mean_anomaly}. Dashed lines show the corresponding relative SVD amplitudes when the snapshot waveforms are generated exactly as in PAPER-I.}
\label{fig:delta_cj_different_eref}
\end{figure}

As in PAPER-I, we find that for binaries with varying $q$, $\chi_1$, $\chi_2$, and $e_{\rm ref}$ the qualitative behavior of the secondary oscillatory term in the SVD phase is unchanged, while its overall amplitude varies. This variation is predominantly linear in $e_{\rm ref}$, and the amplitude also scales approximately linearly with the spherical-harmonic index $\ell$. Together, these trends suggest a universal form for the secondary SVD phase:
\begin{equation}
\frac{\phi_{\ell m j}^{\rm secondary}(\zeta_{\rm ref})}{\ell} = -1.744\,e_{\rm ref}\,\sin(\zeta_{\rm ref}).
\label{eq:secondary_phase_model}
\end{equation}
In Figure~\ref{fig:ph_clmj_secondary}, we show the scaled secondary phase computed from the $(2,2)$, $(3,3)$, and $(4,4)$ modes for four binaries with varying $q$, $\chi_1$, $\chi_2$, and $e_{\rm ref}$. 

These binaries are characterized by $[q,\chi_1,\chi_2,e_{\rm ref}]$ of $[1.5,0.0,0.0,0.1]$, $[2,0.6,0.6,0.15]$, $[2.5,0.2,-0.3,0.12]$ and $[3,0.7,0,0.08]$.
Although the amplitude of $\phi_{\ell m j}^{\rm secondary}$ depends on eccentricity and mode indices, the scaled quantity exhibits a degree of universality. Note that the fit coefficient $1.744$ in Eq.~\eqref{eq:secondary_phase_model} is approximately half the $\ell=2$ coefficient reported in PAPER-I. This $\ell$-dependence of the SVD phase as a function of the relativistic anomaly further connects the eccentric corrections to the $\ell$ and $\ell/2$ scalings observed in recent studies of eccentricity—particularly Refs.~\cite{Islam:2024rhm,Islam:2024bza}—and in the previous section.

Next, following PAPER-I, we inspect the dependence of eccentric-harmonic amplitudes on the relativistic anomaly while holding other binary parameters fixed. We define the relative SVD amplitudes as
\begin{equation}
\frac{\Delta |C_{\ell m j}|}{|C_{\ell m j,0}|}
=\frac{|C_{\ell m j}(\zeta_{\rm ref})|-|C_{\ell m j}(\zeta_{\rm ref}=0)|}{|C_{\ell m j}(\zeta_{\rm ref}=0)|}.
\end{equation}
As found in PAPER-I for the $(2,2)$ mode, the amplitude of the $j=2$ eccentric harmonic (i.e.\ the SVD coefficient magnitude) changes negligibly ($<1\%$) with $\zeta_{\rm ref}$. We confirm that this behaviour also holds for the leading eccentric harmonic in other spherical-harmonic modes. Thus, the amplitude of the leading harmonic indexed by $(\ell,m,j=\ell)$ can be approximated as constant under variations of $\zeta_{\rm ref}$ when all other parameters are fixed. For subleading harmonics, the amplitudes vary significantly and exhibit a periodic dependence that differs from that reported in PAPER-I (which used the \texttt{TEOBResumS} model). We show these relative SVD amplitudes from the $(2,2)$ mode for a binary characterized by $[q,\chi_1,\chi_2,e_{\rm ref}]=[2,0,0,0.1]$ in Figure~\ref{fig:delta_cj_different_eref} (solid lines). We choose a non-spinning eccentric binary to enable a direct comparison with PAPER-I.

\begin{figure}
\includegraphics[width=\columnwidth]{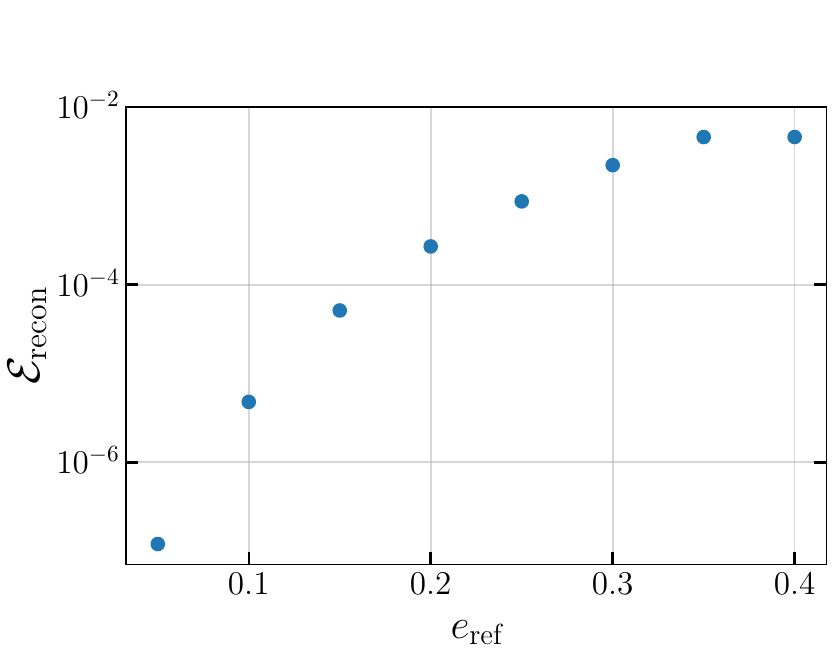}
\caption{We show the reconstruction errors for the $(2,2)$ mode using 50 waveforms in the SVD-based extraction of the first four eccentric harmonics. All waveforms are characterized by $[q,\chi_1,\chi_2]=[2,0.4,-0.2]$ (with different $\zeta_{\rm ref}$ values) and varying $e_{\rm ref}$. We find that the reconstruction error increases with eccentricity. Details are provided in Section~\ref{sec:discussion}.}
\label{fig:recon_error_vs_ecc}
\end{figure}

\section{Discussion}
\label{sec:discussion}
An important question in extracting eccentric harmonics is the choice of reference point. In PAPER-I, we chose a reference time before merger at which the eccentricity and (mean) anomaly were defined. In Section~\ref{sec:extraction}, however, we adopted a slightly different prescription: we fix a common initial orbital frequency as the reference. Consequently, waveforms generated with different (relativistic) anomalies have slightly different lengths. Both choices are valid and may lead to modest differences. We find that the phases (i.e., the arguments of the SVD coefficients across anomaly values) are similar, whereas the coefficient amplitudes can differ visibly. Using the PAPER-I method, we extract the eccentric harmonics from the $(2,2)$ mode for a binary with $[q,\chi_1,\chi_2,e_{\rm ref}]=[2,0,0,0.1]$ and plot the relative SVD amplitudes in Figure~\ref{fig:delta_cj_different_eref} as dashed lines, alongside the results from Section~\ref{sec:model_mean_anomaly} (solid lines). For the dominant harmonic ($j=2$), both approaches agree closely; for subdominant harmonics, the relative SVD amplitudes can differ appreciably.

We also investigated how well our extraction performs as the binary eccentricity increases. Specifically, we vary $e_{\rm ref}$ from $0.05$ to $0.4$ (defined at $t=-30{,}000\,M$ before merger) for $[q,\chi_1,\chi_2]=[2,0.4,-0.2]$. For each case, we use 50 waveforms with different $\zeta_{\rm ref}$ values to obtain the first four eccentric harmonics. We find that the reconstruction error (measured by the relative $L_2$ norm) worsens with eccentricity (cf.\ Fig.~\ref{fig:recon_error_vs_ecc}). We observe similar behaviour for other binaries. This degradation arises because, as eccentricity grows, the time–frequency tracks increasingly overlap (Fig.~\ref{fig:spectrogram}), complicating harmonic separation. Consequently, our framework is most reliable for moderately eccentric binaries ($e_{\rm ref}\lesssim 0.4$). We find similar results for other binaries.

\section{Final remarks}
\label{sec:final_remarks}
In this paper, we have used \texttt{gwMiner} to extract the monotonic eccentric harmonics from the oscillatory spherical-harmonic modes of spinning eccentric binaries, following the methods of  PAPER-I and Ref.~\cite{Patterson:2024vbo}. We then presented the phenomenology of these eccentric harmonics and outlined the observed phase and frequency relations.

One particular aspect of our \texttt{gwMiner} framework is its modularity, allowing extension to other eccentric waveform models. For instance, while PAPER-I employs the \texttt{TEOBResumS} model, this paper uses \texttt{SEOBNRv5EHM} to extract eccentric harmonics. 
Although all these models rely on PN approximations, their internal definitions of eccentricity may differ. In this work, we adopt each model’s internal eccentricity definition, since our focus is on the phenomenology and extraction, which are independent of the definition choice.

It is also important to note that extracting the harmonics for a given mode typically requires $\sim200\,$s. Therefore, constructing a fast approximation—such as a reduced-order model for these eccentric harmonics—is essential for routine data analysis.

Together with PAPER-I, this paper demonstrates the effectiveness of extracting eccentric harmonics for non-precessing waveforms. The next step will obviously be to perform the same analysis for precessing, eccentric binaries. One complication is the lack of sufficient NR simulations and waveform models. Although \texttt{TEOBResumS-DALI} includes both eccentricity and precession effects, the scarcity of NR simulations makes it difficult to assess the model’s accuracy beforehand. Nonetheless, we plan to extend these methods to eccentric, precessing systems in the near future.

One immediate application of this work is its employment in detection pipelines. Due to their simple monotonic behaviour, it is possible to construct efficient templates for eccentric, non-precessing binaries using the eccentric harmonics extracted in this paper. This direction is currently being explored.

Furthermore, we hope this work---together with Refs. Refs.~\cite{Islam:2025rjl,Islam2025InPrep,Patterson:2024vbo}---will encourage the waveform‐modelling community, and in particular the PN modelling groups, to include eccentric harmonics natively in their models (and make them available). By providing theory-driven eccentric harmonics at least through the inspiral, we can avoid reliance on data-driven frameworks and achieve cleaner, cheaper waveforms. 

\begin{acknowledgments}
We thank Steve Fairhurst, Ben Patterson, P. Ajith and Chandra Kant Mishra for useful discussions and comments.
This research was supported in part by the National Science Foundation under Grant No. NSF PHY-2309135 and the Simons Foundation (216179, LB). T.I. is supported in part by the Gordon and Betty Moore Foundation Grant No. GBMF7392 to the KITP.
Use was made of computational facilities purchased with funds from the National Science Foundation (CNS-1725797) and administered by the Center for Scientific Computing (CSC). The CSC is supported by the California NanoSystems Institute and the Materials Research Science and Engineering Center (MRSEC; NSF DMR 2308708) at UC Santa Barbara. 
JR acknowledges support from the Sherman Fairchild Foundation. 
TV acknowledges support from NSF grants 2012086 and 2309360, the Alfred P. Sloan Foundation through grant number FG-2023-20470, the BSF through award number 2022136, and the Hellman Family Faculty Fellowship.  MZ is supported by NSF 2209991, NSF-BSF 2207583, the Nelson Center for Collaborative Research and the Simons Foundation (Simons Collaboration on Black Holes and Strong Gravity). 
\end{acknowledgments}

\bibliography{References}


\end{document}